\documentclass{emulateapj}

\usepackage{natbib}

\newcommand{\dfrac}[2]{{\displaystyle \frac{#1}{#2}}  }

\newcommand{\eqref}[1]{(\ref{#1})}

\def\lesssim{\mathrel{\hbox{\rlap{\hbox{\lower4pt\hbox{$\sim$}}}\hbox{$<$}}}}
\def\gtrsim{\mathrel{\hbox{\rlap{\hbox{\lower4pt\hbox{$\sim$}}}\hbox{$>$}}}}

\slugcomment{ApJ accepted}

\shorttitle{Self-Gravitating Disk with Gap}
\shortauthors{Muto}

\begin{document}

\title{The Structure of a Self-Gravitating Protoplanetary Disk and its
Implications to Direct Imaging Observations}

\author{Takayuki Muto\altaffilmark{1}}
\affil{Department of Earth and Planetary Sciences, 
Tokyo Institute of Technology, \\
2-12-1 Oh-okayama, Meguro-ku, Tokyo, 152-8551, Japan}

\email{muto@geo.titech.ac.jp}
\altaffiltext{1}{JSPS Research Fellow}

\begin{abstract}
We consider the effects of self-gravity on the hydrostatic balance 
 in the vertical direction of a gaseous disk 
 and discuss the possible signature of
 the self-gravity that may be captured 
 by the direct imaging observations of protoplanetary disks in future.   
 In this paper, we consider a vertically isothermal disk 
 in order to isolate the effects of self-gravity.  
 The specific disk model we consider in this paper is the one with a
 radial surface density gap, at which the 
 Toomre's $Q$-parameter of the disk varies rapidly in the radial
 direction.  We calculate the vertical structure of the disk including
 the effects of self-gravity.
 We then calculate the scattered light and the dust thermal emission.  
 We find that if the disk is massive enough and the effects of
 self-gravity come into play,  a weak bump-like structure at the gap edge
 appears in the near-infrared (NIR) scattered light, 
 while no such bump-like structure is seen in the sub-mm dust continuum
 image.  
 The appearance of the bump is caused by the variation of the height of
 the surface in the NIR wavelength.  
 If such bump-like feature is detected in future direct imaging
 observations, with the combination of sub-mm observations,
 it will bring us useful information about the physical states of the 
 disk. 
\end{abstract}

\keywords{protoplanetary disks}

\section{Introduction} 
\label{sec:intro}

Direct imaging observations of protoplanetary disks 
have been making rapid progress.  
In the near-infrared (NIR) wavelength, the strategic observation 
program using Subaru telescope, 
Subaru Strategic Exploration of Exoplanets and Disks (SEEDS), 
is now under way \citep{TamSEEDS}.  
Several results to date reveal that disks have diverse structures
such as an inner hole, a gap or a ring 
\citep{Thalmann10,Hashimoto10}.  
In the submillimeter (sub-mm) wavelength, 
Atacama Large Millimeter/submillimeter Array (ALMA) 
is expected to observe disks at a spatial resolution 
that is comparable to, or better than, NIR wavelength.  

Direct imaging observation is useful in probing the detailed 
structures in the outer disk.  For example, 
a telescope with $0.1$ arc-second resolution has an ability
to resolve the structure with $\sim 10{\mathrm{AU}}$ scale if the disk
is located at $100\mathrm{pc}$.  
Such a spatial resolution is interesting since $10\mathrm{AU}$ is 
comparable with the disk scale height at $100\mathrm{AU}$ away from the
central star.  The disk scale height is an important 
scale in considering the dynamical processes of disks.

From a theoretical point of view, it is important to make a disk model
that can be used for the interpretation of direct imaging observations
of protoplanetary disks, and discuss what we can derive from the
observed structures.  Comparing the information from different
wavelengths is essentially important in order to reveal the real
structure of the observed disks.  

In this paper, we investigate the effects of self-gravity 
on the disk vertical structure in the hydrostatic equilibrium 
and discuss how the signature of self-gravity appears in the 
direct imaging observations.  
We consider the observations of a disk at NIR wavelength, which
mainly observe the scattered light from the upper layer of the disk, 
and the sub-mm dust continuum observations, 
which reveal the surface density structure of the disk.  

One of the fundamental and important properties of the protoplanetary
disk is its mass.  
The sub-mm dust continuum observation is widely used to estimate the
mass of the disk, 
although it suffers the ambiguity of the dust opacity, which
results in at least a factor of two uncertainty in the disk mass.  
In this paper, we consider the effects of self-gravity on the
hydrostatic balance of the disk vertical structure, and discuss how the
effects of self-gravity may appear in the direct imaging observations.
If there is a feature that reflects the self-gravity, it would give
an alternative implication about the disk mass other than the strength
of the dust continuum.   

In order to isolate the effects of self-gravity in the disk structure, 
we consider a disk with a simple radial surface density variation,
namely, a power-law profile with an axisymmetric gap.  
The variation of the surface density may be produced by the
gravitational perturbation by an embedded planet, 
turbulence, or other causes, but we 
do not query the cause of such a structure.  
The question we address is that how such surface density
variations may be observed by direct imaging observations at 
NIR or sub-mm wavelengths and we look for a feature 
that is unique to a self-gravitating disk.  

There are several works on the modeling of 
a self-gravitating disk for direct
imaging observations, especially
in relation to the gravitoturbulence and the resulting planet
formation.  \citet{Narayanan06} have performed a detailed calculations
of line emission based on the hydrodynamic simulations of gravitational
instability by \citet{Boss01}, and have shown that
unique features in $\mathrm{HCO}^{+}$ line may appear close to a forming
giant planet.  \citet{JCB07} have calculated the scattered light using
the same results of hydrodynamic calculations as \citet{Narayanan06}. 
They have shown that the vertically integrated surface density and the
scattered light have no correlation, since strong turbulence induced by a
gravitational instability produces blobs in the upper layer of the disk,
and the scattered light is strongly affected by them.

In this paper, we consider a laminar disk where the vertical hydrostatic
balance is achieved.  Therefore, our model is not useful in a strongly
turbulent disk where the Toomre's $Q$-parameter (defined in the later
section) becomes smaller than $\sim1.5-2$, when non-axisymmetric
gravitational instability causes the disk to be in a turbulent state
(e.g., \citet{Mejia05,Boley06}).  

The paper is constructed as follows.  In Section \ref{sec:model}, we
describe the disk model and the methods of calculations.  In Section
\ref{sec:obsimage}, we show the results of calculations.  In this
section, we also describe the features that appear in the NIR scattered
light image and are unique to self-gravitating disks.  In Section
\ref{sec:diskmass}, 
we discuss the observational implications.  
Section \ref{sec:discuss} is for
summary and further discussions.

\section{Disk Model}
\label{sec:model}

In this section, we describe the basic setup of our model. 
We consider a model where the outer disk has a gapped structure.  
We consider a cylindrical coordinate system $(r,\phi,z)$ with the origin
at the central star.  
As a simple model that features the effects of self-gravity, 
we assume an axisymmetric disk with a gap:
\begin{equation}
 \Sigma(r) = 2.7 f_{\rm enhance} f_{\rm gap}(r)
  \left(\dfrac{r}{100\mathrm{AU}}\right)^{-1.5} \mathrm{g/cm}^2,
  \label{surf_model}
\end{equation} 
where $f_{\rm enhance}$ is a parameter that controls the overall 
surface density of the disk and $f_{\rm gap}(r)$ is a function that 
determines the shape of the gap.  
We vary the disk mass by changing $f_{\rm enhance}$ from $1$ to $8.15$, 
but in the subsequent section, we especially consider the case of
$f_{\rm enhance}=1$ and $6.6$ as illustrative examples.  The case of
$f_{\rm enhance}=6.6$ is rather extreme, since the typical $Q$-value in
this case is close to $2$, where turbulence may take place.  However,
the effects of self-gravity on the vertical structure is prominent in
this extreme case.  
We assume a Gaussian gap profile,
\begin{equation}
 f_{\rm gap}(r) = 1 - \alpha 
  \exp \left[ 
	-\left( \dfrac{r-r_{\rm gap}}{\Delta_{\rm gap}} \right)^2
       \right], 
  \label{gap_model}
\end{equation} 
with $\alpha=0.95$, $r_{\rm gap}=100\mathrm{AU}$, 
and $\Delta_{\rm gap}=30\mathrm{AU}$.  The parameters $\alpha$, 
$r_{\rm gap}$, and $\Delta_{\rm gap}$ parameterize the depth, location
and width of the gap, respectively.  

In this paper, in order to make the problem simple and tractable, 
we make a simple assumption for the temperature of the disk.
We assume a vertically isothermal disk where the temperature $T(r)$ is
given by
\begin{equation}
 T(r) = 30 \left( \dfrac{r}{100\mathrm{AU}}  \right)^{-1/2} \mathrm{K}.
  \label{temp_model}
\end{equation}
Although we do not question ourselves about the origin of 
such surface density and temperature structure in detail, we comment
here that it is reported that a massive planet emedded in a disk causes
a deep gap in the surface density profile, while the 
temperature does not vary as significantly as the surface density
\citep{DAngelo03}. 

\begin{figure}
 \plottwo{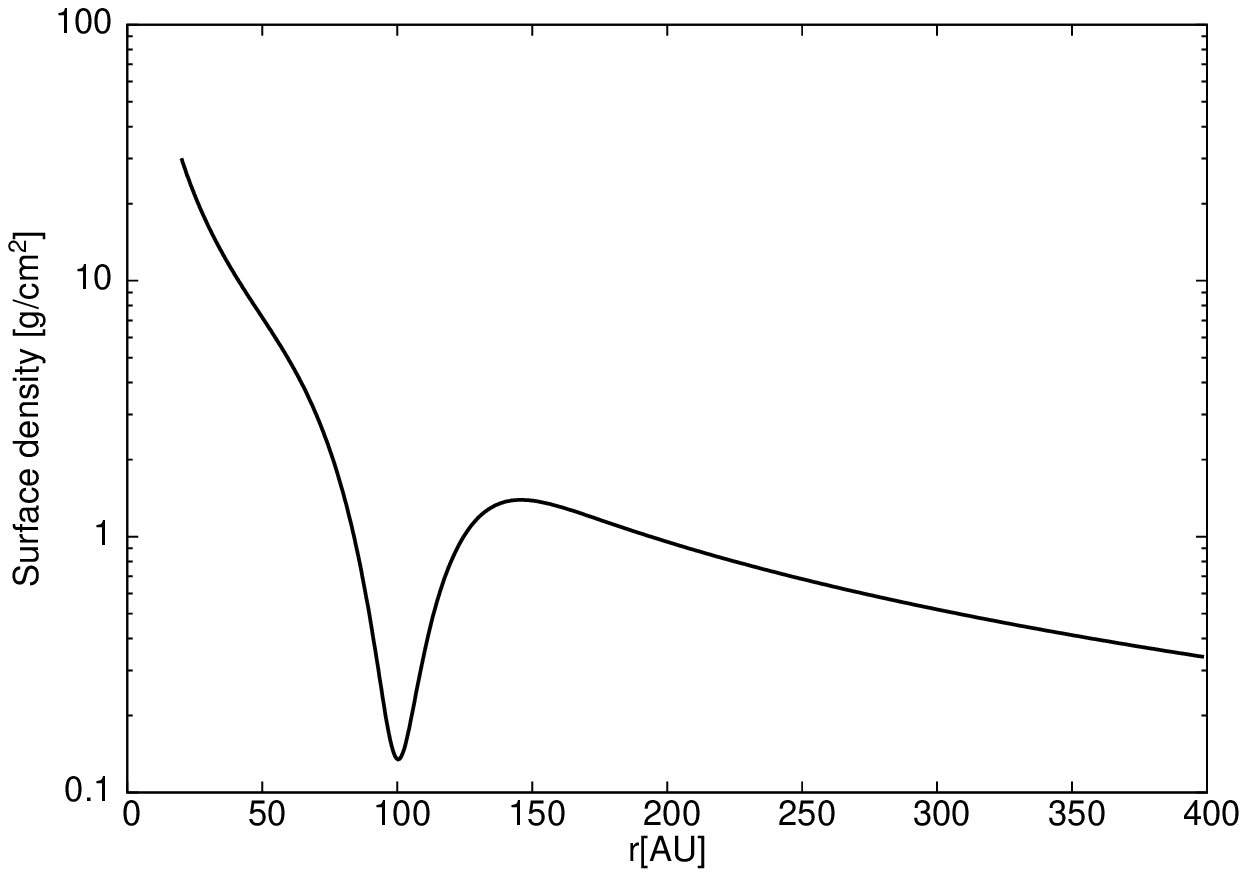}{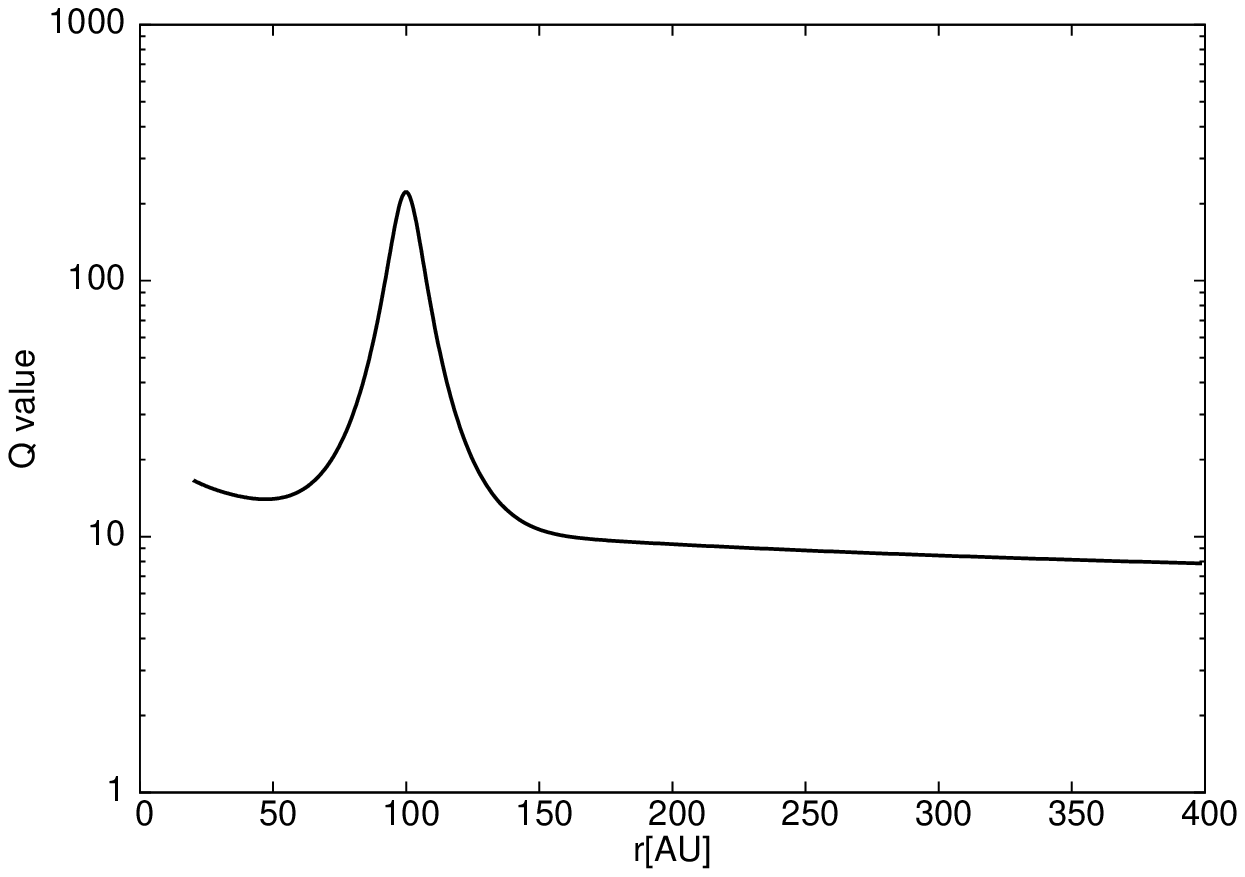}
 \caption{ An example of the surface density profile and the disk
 $Q$-parameter profile in the model.  
 The left panel shows the surface density of the gas with
 $f_{\rm enhance}=1$ in equation \eqref{surf_model}, and the right panel
 shows the disk $Q$-parameter for this model. 
 }
 \label{fig:diskmodel}
\end{figure}

By varying the overall surface density by $f_{\rm enhance}$ while
keeping the temperature profile fixed, 
we vary the disk Toomre's $Q$-parameter, 
$Q = \Omega c/\pi G \Sigma$.  The $Q$-parameter is also varied in the
radial direction due to the gap.  Figure \ref{fig:diskmodel} shows one
example of the disk model.  In this figure, we show the profile of 
surface density and $Q$-value with $f_{\rm enhance}=1$.

From the surface density and the temperature profile, 
we obtain the vertical structure of the disk. 
We assume that the hydrostatic equilibrium is reached 
and that the disk is geometrically thin.  We solve
\begin{equation}
 \dfrac{dp(z)}{dz} = -\rho(z) \left( \Omega^2 z +
			       \dfrac{\partial\psi_{\rm d}}{\partial z}
			      \right),
 \label{hydro_eq}
\end{equation}
where $p=c^2 \rho$ is the pressure and $\Omega(r)=(GM_{\ast}/r^3)^{1/2}$
is the Keplerian angular frequency.  
The sound speed $c(r)$ is obtained from the temperature profile given by
equation \eqref{temp_model} by
\begin{equation}
 c(r) = 10^{5} \left( \dfrac{T(r)}{300\mathrm{K}} \right)^{1/2} 
  \mathrm{cm/s}.
\end{equation}
In equation \eqref{hydro_eq}, $\psi_{\rm d}$ denotes the gravitational
potential of the disk, which is determined by Poisson equation
\begin{equation}
 \nabla^2 \psi_{\rm d} = 4 \pi G \rho.
  \label{poisson}
\end{equation}
We solve equations \eqref{hydro_eq} and \eqref{poisson} iteratively to
find a consistent solution.  If the variation of the surface density is
slow in the radial direction, the local approach by \citet{Pac78}
yields a reasonable model.  However, in the model presented in this
paper, such variation may be rapid due to the presence of the gap, 
so it is necessary to solve full Poisson equation.  

Once we obtain the density structure of the disk, we solve the equations
of radiative transfer to obtain the observed image.  
We consider the thermal emission of the disk 
and the light from the central star scattered by the disk
surface.  The former is important in the observations in sub-mm
wavelength and the latter is important in the NIR imaging observations.  
For simplicity, we assume that 
the dust and gas are well-mixed and the optical
properties of the dust particles do not vary in the disk.
We also assume that the disk is face-on.  

We follow the approach by \citet{DAlessio99} in calculating the thermal
emission and the scattered light.  
The thermal emission may be calculated by integrating
\begin{equation}
 \dfrac{dI^{\rm therm}}{dZ} = - \kappa \rho B(T) \exp
  \left[ -\tau(Z) \right],
\end{equation}
where $Z$ is the coordinate along the line of sight, $\kappa$ is the
absorption cross section per unit gas mass, 
$B(T)$ is the Planck function, and $\tau$ is the optical depth along the
line of sight, which is calculated by 
\begin{equation}
 \dfrac{d\tau}{dZ} = - \chi \rho,
\end{equation} 
where $\chi$ is the total extinction (including absorption and
scattering) cross section per unit gas mass.  
In calculating the scattered light, we assume a single and isotropic
scattering.  Then, the scattered light can be calculated by integrating 
\begin{equation}
 \dfrac{dI^{\rm scat}}{dZ} = -\sigma W(r,z) B(T_{\ast}) 
  \exp\left[ -\tau_{\rm rad} - \tau(Z) \right],
  \label{transfer_scat}
\end{equation} 
along the line of sight.  Here, $\sigma=(\chi-\kappa)\rho$ 
is the scattering coefficient,
$W(r,z)=1/4(r^2+z^2)$ ($r$ is the radial coordinate and $z$ is the
vertical coordinate) is the geometric dilution
factor \citep[e.g.,][]{JC09}, 
$B(T_{\ast})$ is the blackbody
radiation of the central star, and $\tau_{\rm rad}$ is the optical depth
for the stellar irradiation.  
In determining $W(r)$, we assume that the central star is a point
source, which is a reasonable assumption when considering the outer
disk.

The opacity of dust particles is one of the major uncertainties in the
observational modeling of the disk.  Therefore, we use a range of values
of the opacity in order to find an opacity-independent signature.  
We use the total extinction $\chi$ 
with $40-200\mathrm{cm^2/g}$ for the modeling of the NIR observation, 
and $1.2\times10^{-4}-6\times 10^{-3}\mathrm{cm^2/g}$
for the modeling of sub-mm observation.  
In the subsequent section, where we describe the effects of self-gravity
in detail, we show the results of $\chi=100\mathrm{cm^2/g}$ for NIR
observations and $\chi=0.003\mathrm{cm^2/g}$ for sub-mm observations.  
We use the fixed values of
scattering albedo, $\omega = (\chi-\kappa)/\chi$ for simplicity.  
We use $\omega=0.9$ for the NIR modeling and $\omega=10^{-3}$ for
the sub-mm modeling.

\section{Signature of a Self-Gravitating Disk}
\label{sec:obsimage}

\begin{figure}
 \plottwo{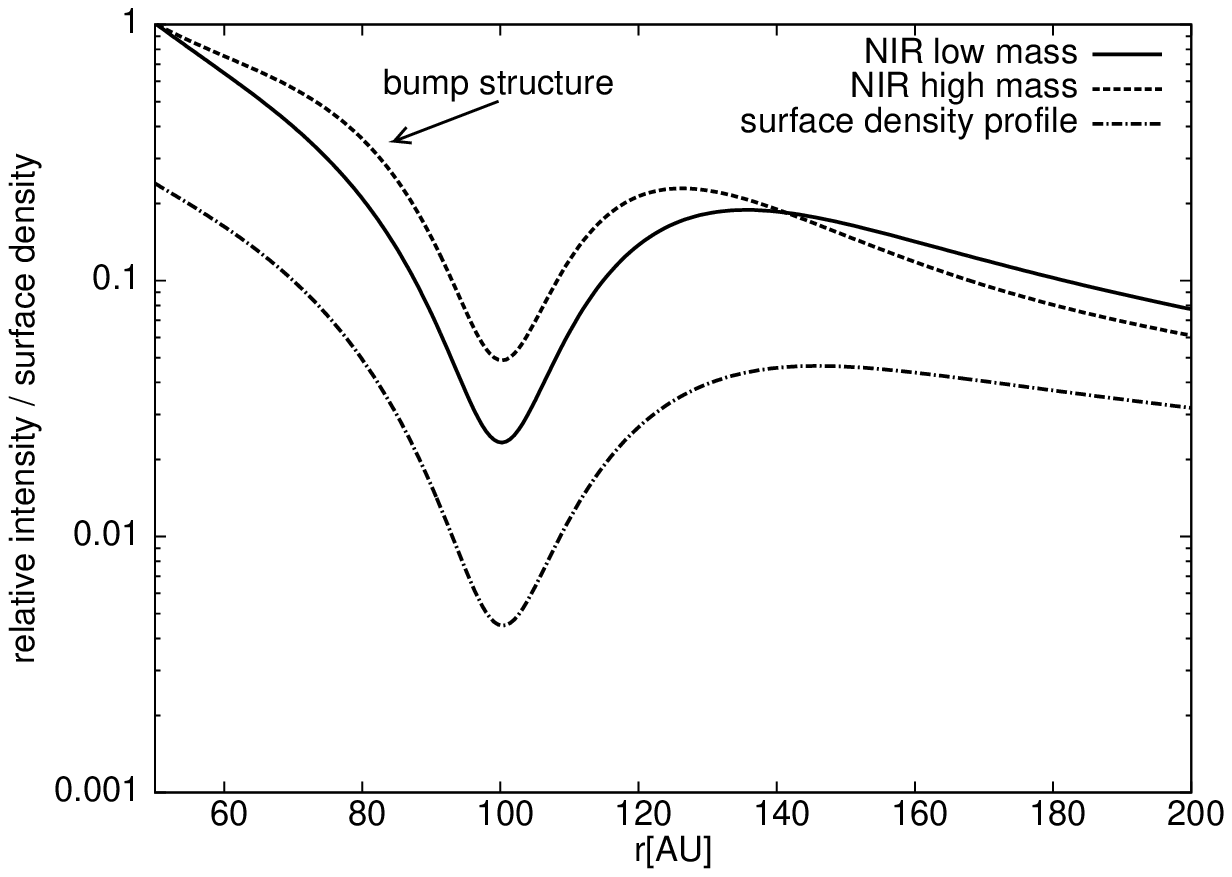}{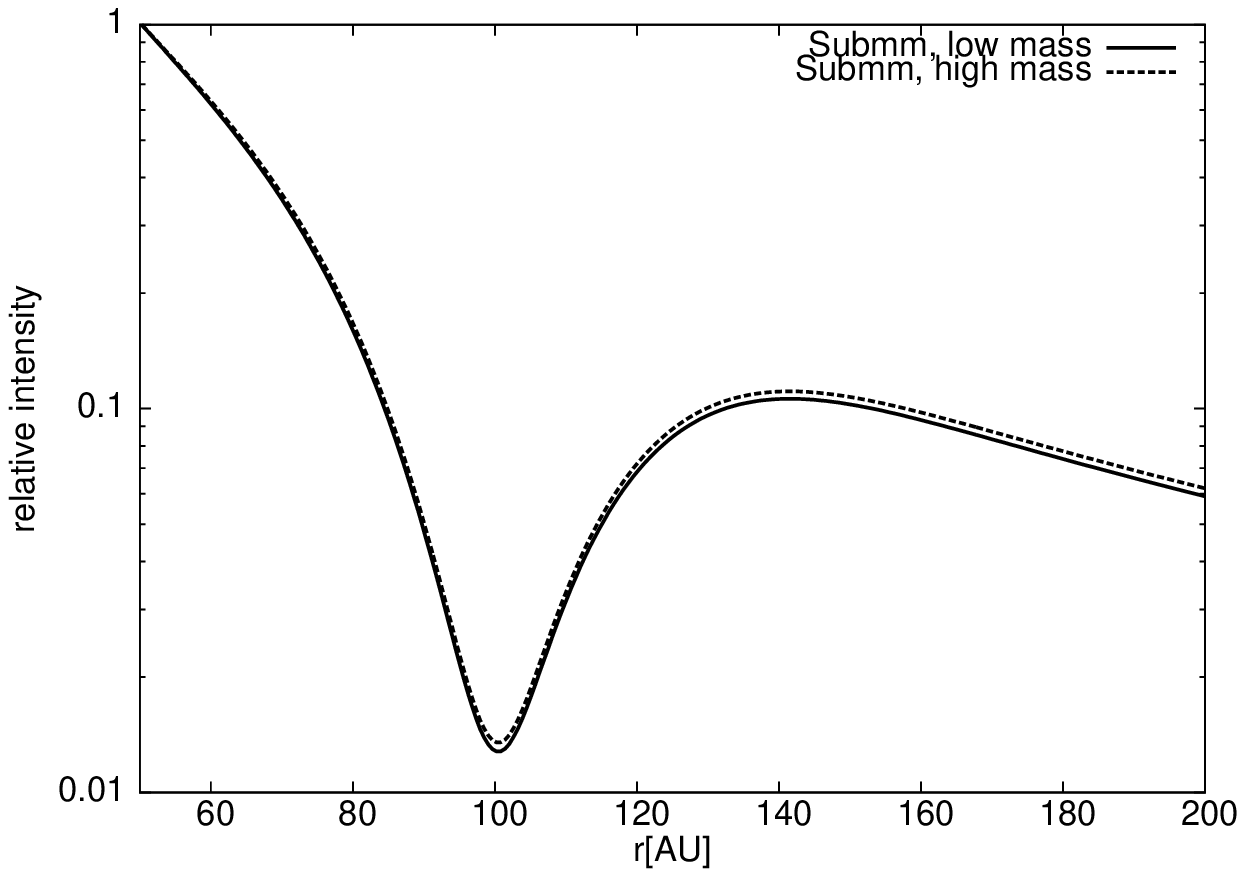}
 \caption{ The radial profile of NIR observation (left) and sub-mm
 (right) observation.  The surface brightness normalized by that at
 $50\mathrm{AU}$ is shown.  The solid line corresponds to the disk model
 with $f_{\rm enhance}=1$ (low-mass model) and the dashed line is for
 $f_{\rm enhance}=6.6$ (high-mass model).  The results with extinction
 $\chi=100\mathrm{cm}^2\mathrm{/g}$ (NIR) and 
 $\chi=0.003\mathrm{cm}^2\mathrm{/g}$ (sub-mm) is shown.
 In the left panel, we also show the profile of the surface density in
 the arbitrary unit by the dot-dashed line.  The arrow indicates the
 ``bump'' structure which is mainly discussed in this paper.
 }
 \label{fig:relI}
\end{figure}

Since we assume that the disk structure is axisymmetric and the disk is
face-on, the resulting image is also axisymmetric.  
Figure \ref{fig:relI} shows the examples of the 
observed radial profile of the relative 
intensity in NIR wavelength and sub-mm wavelength.  
The disk brightness is normalized by that at $50\mathrm{AU}$.  
The disk with $f_{\rm enhance}=1$ (low-mass disk) and the disk with
$f_{\rm enhance}=6.6$ (high-mass disk) are shown.  
The total extinction coefficient $\chi$ of the disk is 
$100\mathrm{cm}^2\mathrm{/g}$ in NIR
and $0.003\mathrm{cm}^2\mathrm{/g}$ in sub-mm, respectively.

In the sub-mm observation, 
there is no significant difference in the profile of the disk brightness
between the low-mass disk case and the high-mass disk case.  This is
because in the sub-mm wavelength, the disk is optically thin, and the
observed flux is proportional to the disk surface density.  In our
model, the profile of the surface density is the same in the low-mass
and the high-mass case except for the factor $f_{\rm enhance}$.  
Therefore, the relative intensity profiles are
essentially the same in both cases.

In the NIR observation, on the other hand, 
we see a difference in the radial structure of the relative intensity
between the low-mass disk case and the high-mass disk case.  
In the case of the low-mass disk, the profile of the relative intensity
is very similar to that of sub-mm.  There is a dark region in the disk
around the place where the gap is present.  In the case of the high-mass
disk, we see that there is a bump in the relative intensity profile 
at $\sim80\mathrm{AU}$ region (indicated by an arrow) beside the
dark region around $\sim100\mathrm{AU}$.
We also note that there is another bump-like structure outside the gap
region.  In Figure \ref{fig:relI}, we see such a structure around
$120-140\mathrm{AU}$ region.  In the region further away from the
central star, we see that the NIR brightness is weakened.

We now focus on the bump-like structure that appears in the NIR profile
just inside the gap region (indicated by an arrow in Figure
\ref{fig:relI}), and we shall show that this structure indeed
arises from the effects of self-gravity.  

\begin{figure}
 \plottwo{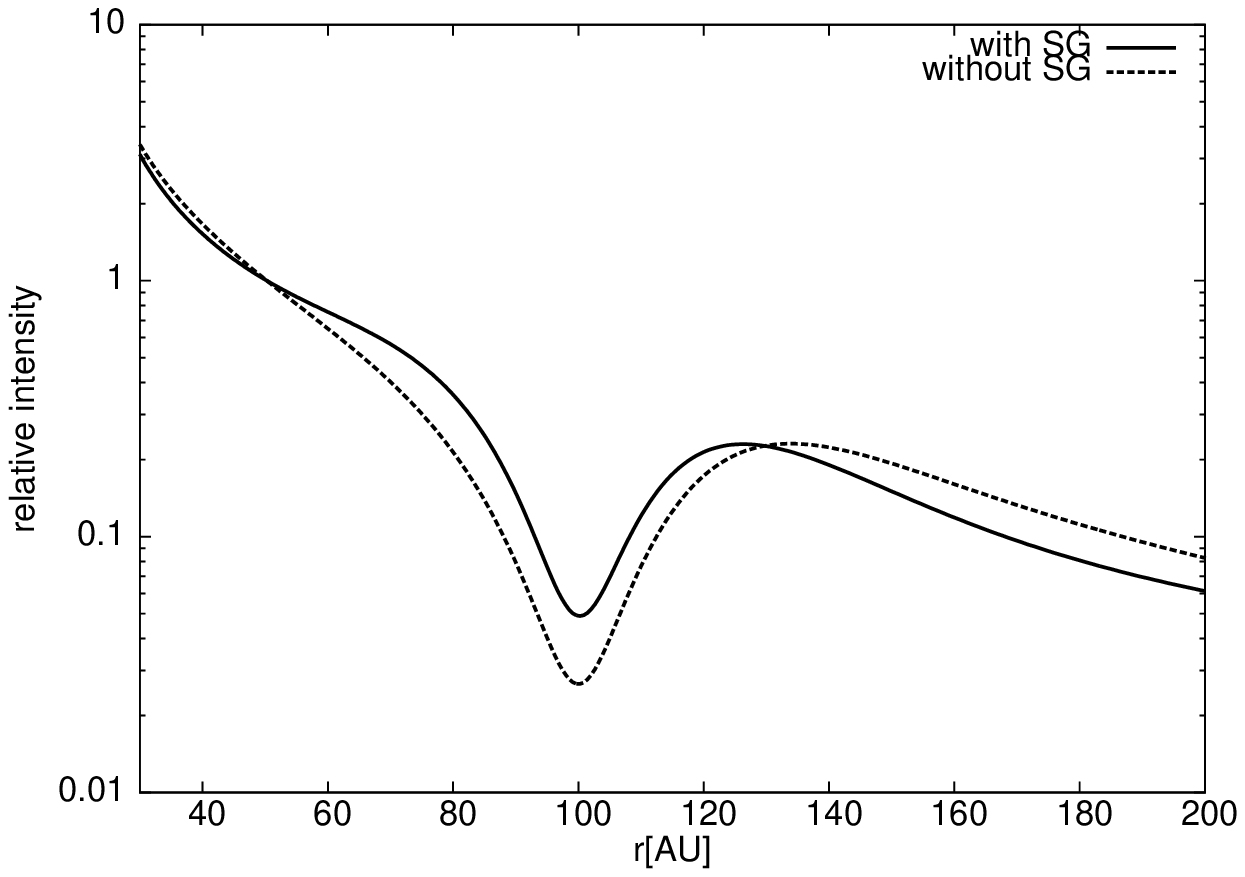}{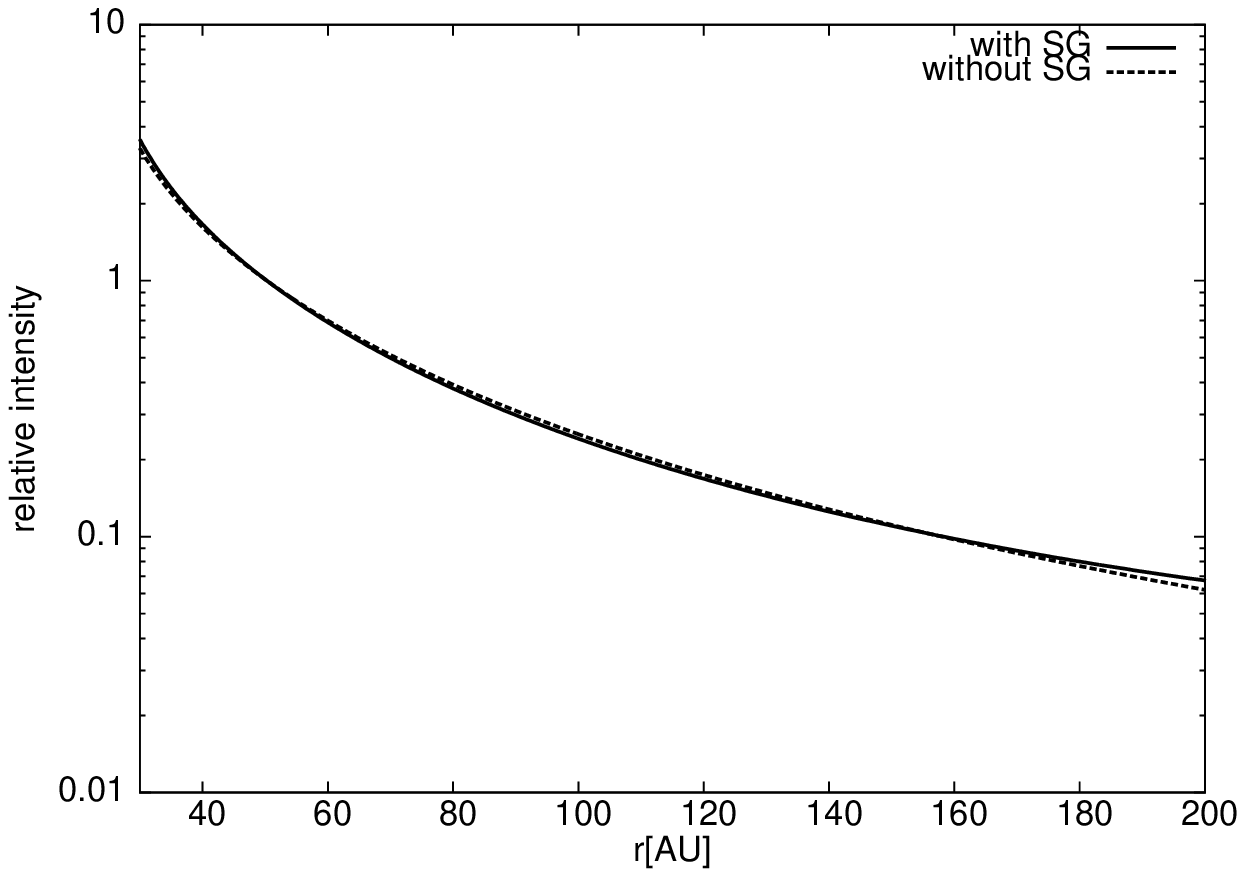}
 \caption{ 
 Comparison between the calculations with self-gravity and without
 self-gravity.  The left panel shows the disk model with a gap, and the
 right panel shows the disk model without a gap.  The solid line shows
 the model with self-gravity, and the dashed line is the model without
 self-gravity.  We use the model with $f_{\rm enhance}=6.6$, which
 corresponds to the ``high-mass model'' in Figure \ref{fig:relI}.  
 }
 \label{fig:relI_SGcomp}
\end{figure}

In the left panel of Figure \ref{fig:relI_SGcomp}, 
we compare the results of NIR scattered light with and without
self-gravity.  We use the high-mass disk model 
with $f_{\rm enhance}=6.6$.   
The solid line shows the same calculation 
with the dashed line in the left panel of
Figure \ref{fig:relI} and 
the dashed line in Figure \ref{fig:relI_SGcomp}
shows the results for the same disk model 
but the self-gravity is artificially turned off.  
We see the bump structure around $\sim 80\mathrm{AU}$ region in
the calculations with self-gravity but it is not significant in
the calculations without self-gravity.  

In order for the bump structure to be seen, it is also necessary to have
a gap in the surface density structure.  In the right panel of Figure
\ref{fig:relI_SGcomp}, we show the relative intensity profile of the NIR
scattered light derived from the calculations with $f_{\rm enhance}=6.6$
but $f_{\rm gap}(r)=1$ everywhere (i.e., no gap).   
We compare the calculations with and without self-gravity, 
but there is no significant difference between the two calculations.  

\begin{figure}
 \plotone{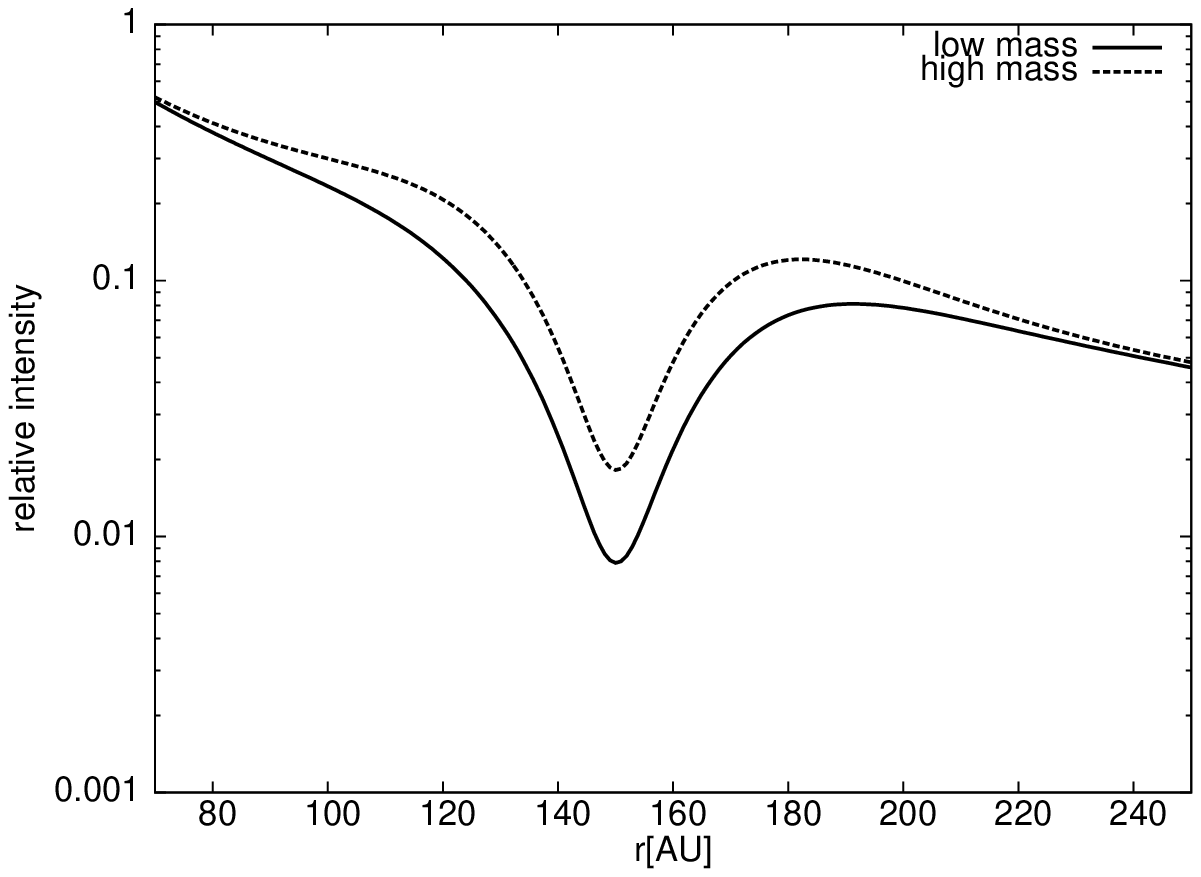}
 \caption{ 
 The results of NIR profile for the model with the gap at
 $150\mathrm{AU}$.  The models with $f_{\rm enhance}=1$ (solid line) and 
 with $f_{\rm enhance}=6.6$ (dashed line) are shown.  
 The bump inside and outside the gap region is seen in the case of the
 high mass disk. 
 }
 \label{fig:NIR_gap150au}
\end{figure}

\begin{figure}
 \plotone{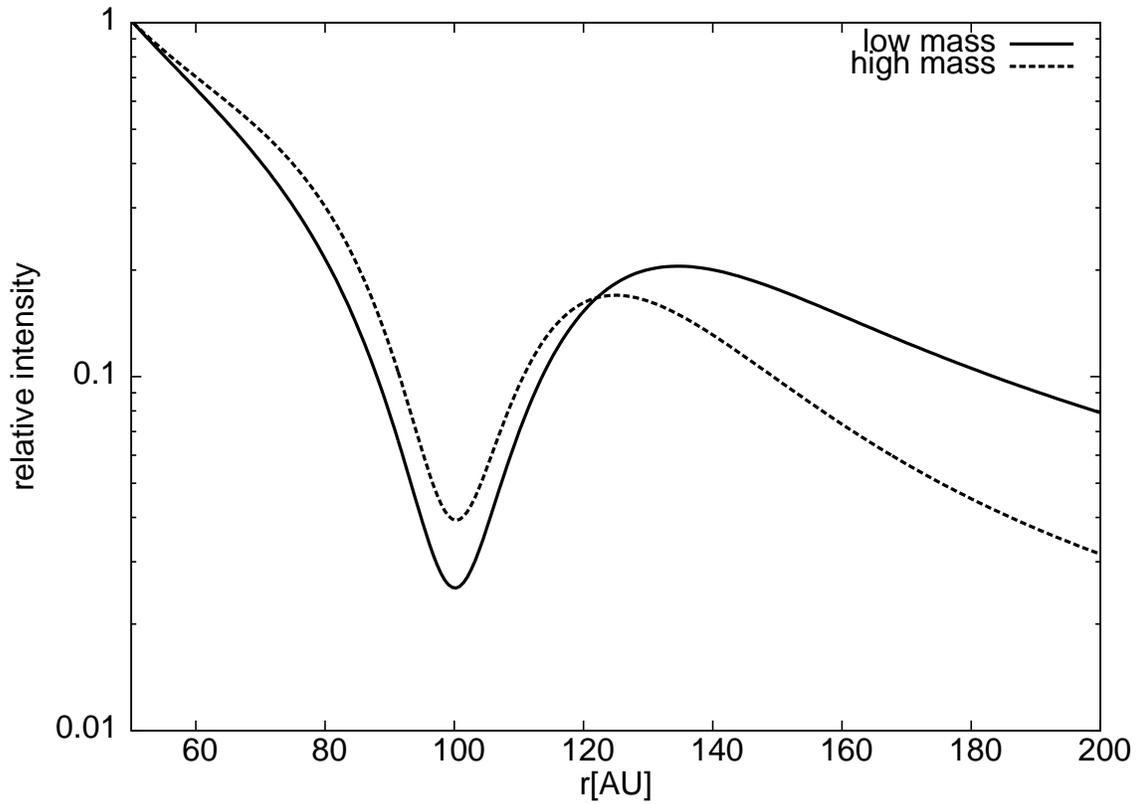}
 \caption{ 
 The results of NIR profile for the model with the surface density
 profile of $\Sigma \propto r^{-1}$.  The models with 
 $f_{\rm enhance}=1$ (solid line) and with $f_{\rm enhance}=6.6$ (dashed
 line) are shown.  
 The bump inside and the outside the gap region is seen in the
 case of the high mass disk.
 }
 \label{fig:NIR_p1}
\end{figure}

As long as the disk is massive enough and there exists a gap structure
in the surface density, the bump structure in the NIR profile is always
present regardless of the location of the gap 
or the underlying surface density structure of the disk.  
Figure \ref{fig:NIR_gap150au} shows the NIR
profile for the model with $r_{\rm gap}=150\mathrm{AU}$.  
In this figure, we compare the results with $f_{\rm enhance}=1$ 
and $f_{\rm enhance}=6.6$.  
We see that there is a bump structure just inside (and also outside) of
the gap region.  
In Figure \ref{fig:NIR_p1}, we show the results of the NIR
profile for the disk model with $\Sigma \propto r^{-1}$.  
The disk model is the same with equations
\eqref{surf_model}-\eqref{temp_model} except for the power of the
surface density distribution, and we assume $\alpha=0.95$, 
$r_{\rm gap}=100\mathrm{AU}$ and $\Delta_{\rm gap}=30\mathrm{AU}$ for
the gap profile.  We compare the
results with $f_{\rm enhance}=1$ and $f_{\rm enhance}=6.6$.  In this
model, again, we see a bump-like structure inside and outside of the gap
region for $f_{\rm enhance}=6.6$.

We now discuss the origin of the bump structure in the NIR observation.  
In the NIR observation, we mainly observe the light from the central
star scattered at the surface of the disk.  We define the
surface of the disk as the place where the optical depth towards the
central star, $\tau_{\rm rad}$, becomes unity.  

The strength of the scattered light, which is given by equation
\eqref{transfer_scat}, can be well approximated by using the grazing
angle, which is the angle between the incident ray from the central star 
and the disk surface~\citep[see e.g., ][]{Inoue08,JC09}.  The grazing
angle can be calculated from the density profile.   
We have checked that the scattered light profile is indeed well
reproduced by the grazing angle times a geometrical factor.  
Therefore, we discuss the effects of self-gravity on the behavior of the
surface of the disk and the grazing angle in the following paragraphs.  
We note that all the scattered light profiles given so far are
calculated by directly integrating equation 
\eqref{transfer_scat}.  

We first consider the case where the self-gravity of the disk is
negligible.  
In Appendix, we show explicit calculations of the disk surface
and the grazing angle in the case where vertical hydrostatic equilibrium
is reached in a non-self-gravitating disk.  
It is shown that (1) the more the disk material, the higher in the
altitude the disk surface is (see equation \eqref{surf_approx}), and 
(2) grazing angle $\beta$ is approximately given by 
\begin{equation}
 \beta \sim h \dfrac{\rho_0}{\langle \rho_0 \rangle},
\end{equation} 
where $h=c/r\Omega$ is the aspect ratio of the disk, $\rho_0$ is the
density at the disk midplane, and $\langle \rho_0 \rangle$ is the radial 
average of the midplane density calculated by 
\begin{equation}
 \langle \rho_0 \rangle 
  = \dfrac{1}{r} \int dr^{\prime} \rho_0 (r^{\prime})
\end{equation}
 (see equation \eqref{beta_approx}).  
If there exists a gap in the radial surface density profile and the
width of the gap is small compared to the disk scale, 
the midplane density at the gap location is low while 
the radially averaged density is not significantly affected by the
presence of the gap.  
Therefore, we conclude that the scattered light traces the
surface density profile well, thereby the NIR scattered light is well
correlated with sub-mm dust continuum emission.
It is possible to see in the left panel of Figure \ref{fig:relI} that
the location of the dark region in the NIR profile and the location
of the gap in the surface density profile are very close each other.  

\begin{figure}
 \plotone{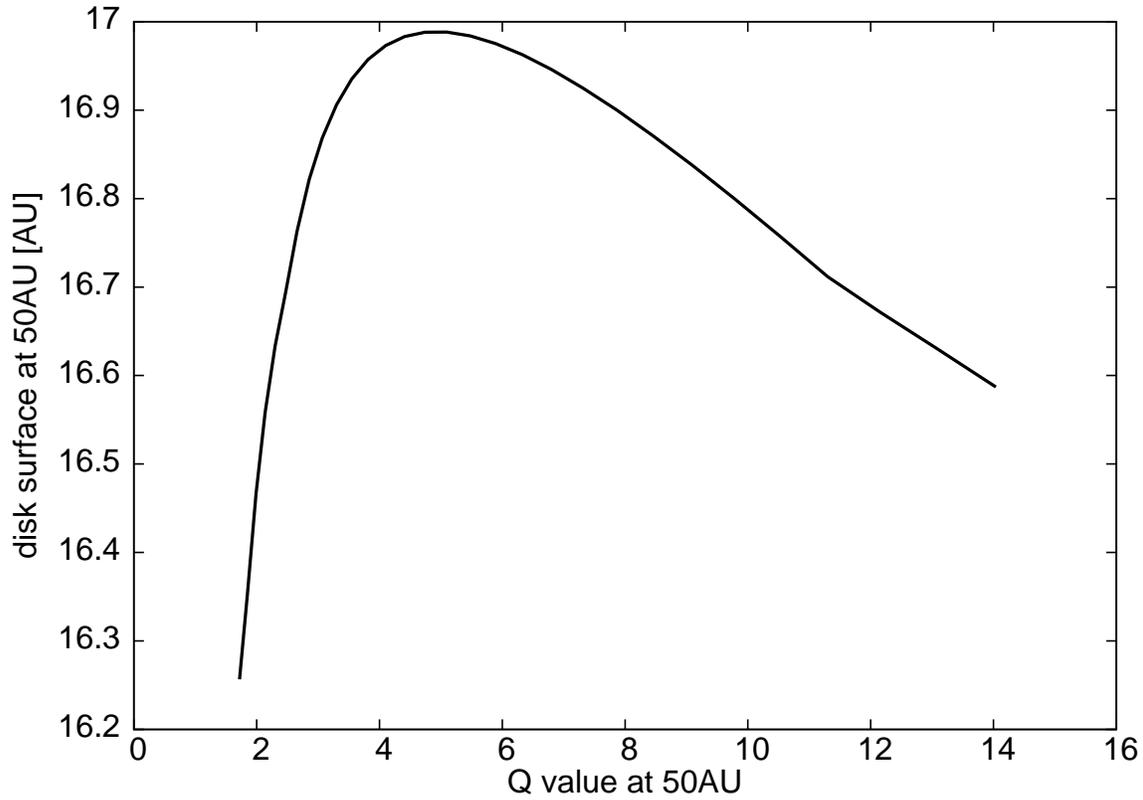}
 \caption{ The height of the disk surface (in AU) for various disk mass.
 The surface at $50\mathrm{AU}$ as a function of the $Q$-parameter at
 $50\mathrm{AU}$ is shown.  The surface is defined as a place where the
 optical depth towards the central star is unity.  The results with
 $\kappa=100\mathrm{cm}^2\mathrm{/g}$ are shown.
 }
 \label{fig:Qsurf}
\end{figure}

If we include the effects of self-gravity, it acts to pull the disk
material towards the disk midplane.  Therefore, the disk self-gravity in
general lowers the disk surface.  
Therefore, we expect that the position of the disk surface is maximized
at a certain value of $Q=Q_c$.  
Figure \ref{fig:Qsurf} shows the disk surface at 
$50\mathrm{AU}$ (it is well away from the gap region) for various
models with $f_{\rm enhance}$.  The horizontal axis shows the disk
$Q$-value at $50\mathrm{AU}$.  
It is shown that the disk surface is 
at the highest place when $Q = Q_c \sim 5$. 
The exact value of $Q_c$ depends on the place of the disk we consider.
For example, we have checked that 
the value of $Q_c$ is approximately $3$ at $200\mathrm{AU}$.
However, the overall behavior of the location of the disk surface as a
function of the disk mass is similar.  

\begin{figure}
 \plotone{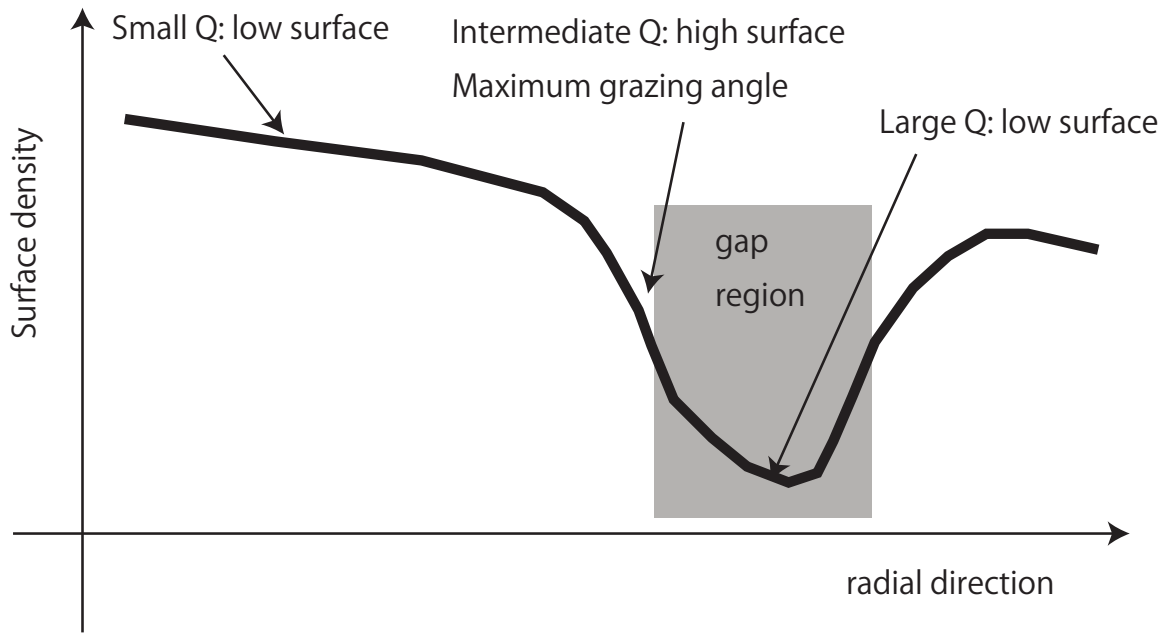}
 \caption{ 
 Schematic diagram for the origin of bump at the gap edge in NIR
 observation.  In the region where the $Q$-value is small, the surface
 is a relatively lower place because of the effects of the
 self-gravity.  In the region of the gap, the surface is also at a 
 lower place because of the lack of the material.  For the intermediate
 values of $Q$, the surface is at the highest place and therefore, the
 grazing angle is maximum, which leads to the formation of bump in the
 scattered light.  
 }
 \label{fig:NIR_schematic}
\end{figure}

\begin{figure}
 \plotone{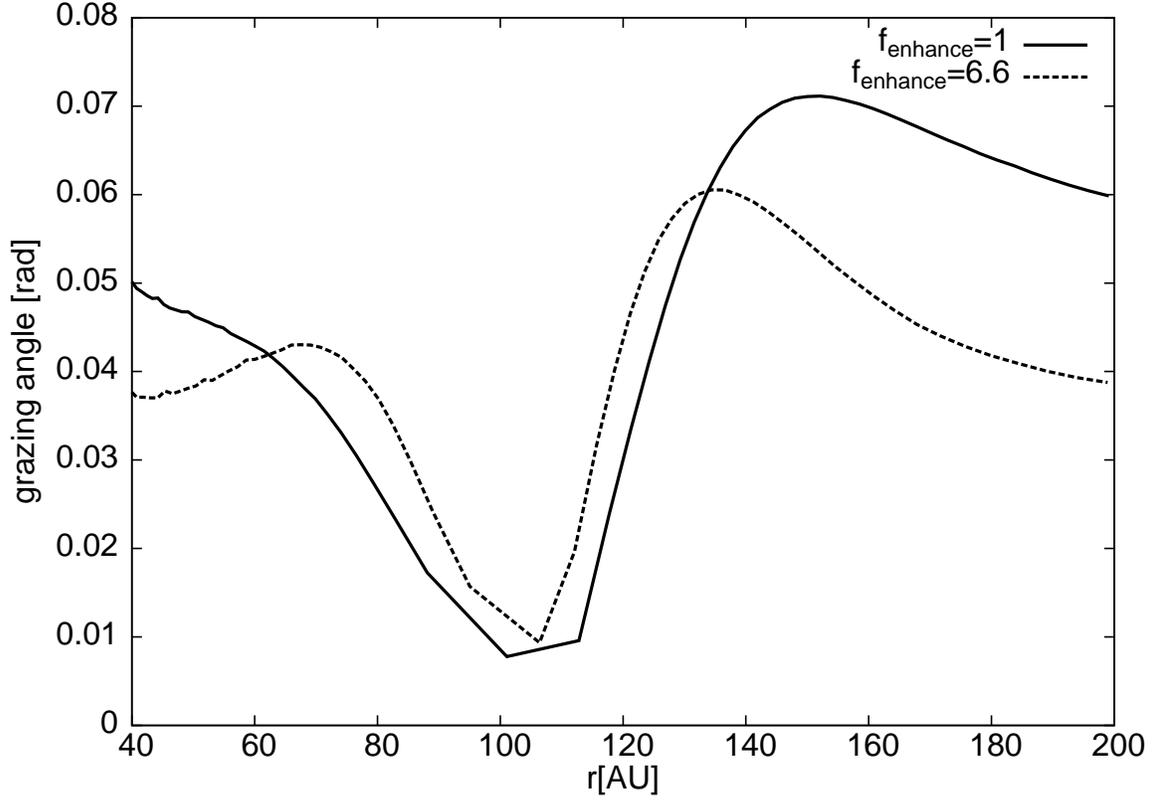}
 \caption{ 
 The grazing angle (in radian) as a function of radial distance (in AU)
 in two different $f_{\rm enhance}$ parameters.  
 The solid line shows the case of $f_{\rm enhance}=1.0$ 
 and the disk self-gravity is almost negligible.  
 The dashed line shows the case of $f_{\rm enhance}=6.6$.
 The effect of self-gravity appears at the gap edge, and there is a
 maximum of the grazing angle around $70\mathrm{AU}$ region.  
 The results with $\kappa=100\mathrm{cm}^2\mathrm{/g}$ are shown.
 }
 \label{fig:graze_angle}
\end{figure}

Let us now consider what Figure \ref{fig:Qsurf} indicates for one
particular disk model, in which there is a rapid variation of
$Q$-parameter in the radial direction.  
From Figure \ref{fig:Qsurf}, it is possible to say that 
if there exists a place where the disk $Q$-parameter changes rapidly and
crosses $Q \sim Q_c$, 
we expect that the disk surface is at the highest
place at such radius.  The grazing angle is maximized there, and
therefore, the scattered light from such radius is prominent.  
Figure \ref{fig:NIR_schematic} shows a schematic picture of this.  
In our disk model, there is a significant change in the disk $Q$-value 
at the location of the gap, 
and at the edge of the gap, there exists a place where $Q \sim Q_c$.  
Figure \ref{fig:graze_angle} compares the profile of the grazing angle
in the case of $f_{\rm enhance}=1$ (almost non-self-gravitating) and
$f_{\rm enhance}=6.6$ (self-gravitating disk).  It is clearly shown that
the grazing angle is maximized at the gap inner edge in the
case where self-gravity becomes important.  

The appearance of the maximum of the grazing angle at the edge of the
gap depends on how massive the disk is and to what extent self-gravity
effects are important in the vertical structure of the disk.  
The bump is more enhanced if the disk is more massive.
In order to see this, we calculate the strength of the peak at the inner
edge of the gap as follows.  
First, we fit the brightness profile
between $35\mathrm{AU}$ and $50\mathrm{AU}$ by power law.  
We have chosen this region from our data 
since this region is not significantly affected by the
presence of the gap (see Figure \ref{fig:relI_SGcomp}).  
Then, we normalize the profile with the power law profile  
obtained.  Finally, we look for the maximum between $60\mathrm{AU}$ and
$100\mathrm{AU}$ from the normalized profile to find the location and
the strength of the peak.  
The choice of $60\mathrm{AU}$ and
$100\mathrm{AU}$ is arbitrary, but the point here is that the bump
in NIR profile is included in this region.  

Figure \ref{fig:NIRpeak_Q} shows the relation between the peak strength
and the averaged $Q$-value between $35\mathrm{AU}$ and $50\mathrm{AU}$.
The models with various values of $f_{\rm enhance}$ and dust extinction
are shown.  
The different points sharing the same $Q$-value 
correspond to the different values of dust extinction at NIR.
It is shown that the smaller the $Q$-value, the stronger the peak, and
that there is only small range of the $Q$-value that can cause the peak
strength of $1.15-1.2$.  There is no significant peak for low-mass
disks.  
For the particular model of $f_{\rm enhance} = 6.6$, which we have
discussed in detail in the previous section, the peak strength is of the
order of $10\%$.

In summary, the bump structure at the edge of the gap 
in the NIR scattered light profile is a characteristic feature arising
from the effects of the self-gravity.  
It is also noted that, in the discussion above, the
essential assumption is the vertical hydrostatic equilibrium, rapid
variation of the $Q$-parameter, and the dominance of the scattered light
in NIR observations.  We therefore expect that the profile of the
surface density is not necessarily a ``gap'' in order for the bump in
the NIR imaging observations to be produced.  
It is not necessarily axisymmetric, and it can be a ``hole'' or other
such features that exhibit the change in the $Q$-parameters, as long as
the hydrostatic equilibrium is reached.

\section{Discussion: NIR Imaging Observations of High-Mass Disks}
\label{sec:diskmass}

We have seen that the bump structure at the gap edge in the NIR
observations reflects the fact that the disk self-gravity 
comes into play in determining the vertical structure of the disk.  
We now discuss the implications of the results to the observations.  

\begin{figure}
 \plotone{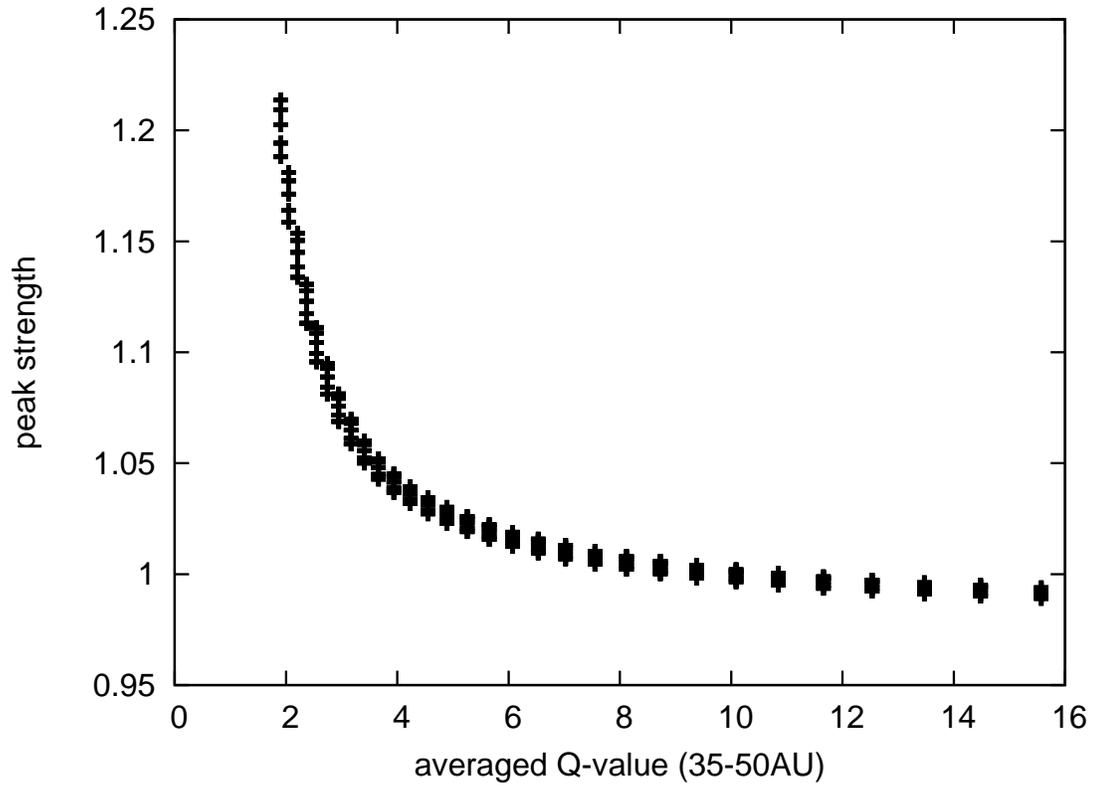}
 \caption{
 The relation between the strength of the peak of the bump in the NIR
 observations and the $Q$-value.  
 The $Q$-values are averaged over the radius between
 $35\mathrm{AU}$ and $50\mathrm{AU}$.  Different points with the same
 values of $Q$ correspond to different extinction coefficients $\chi$
 in the NIR wavelength.
 }
 \label{fig:NIRpeak_Q}
\end{figure}

We first note that the peak strength shown in Figure \ref{fig:NIRpeak_Q}
can, in principle, be calculated from the observed relative intensity
profile only.   
From Figure \ref{fig:NIRpeak_Q}, it may be possible to say that   
if we can detect the peak at the edge of the gap, 
it may be possible to constrain the disk $Q$-value
more effectively than by using the sub-mm continuum emission profile
alone, which always has an uncertainty of at least a factor of two
arising from the uncertainty of the dust parameters.  

It is noted, however, that this argument would apply for the disk with
a limited range of mass.  
For low-mass disks (having high $Q$-value),
the peak is not very significant and it is difficult to constrain the
disk mass.  This method of estimating the disk mass 
can be useful for the disk with the $Q$-value less than 4 or around.
It is also noted that for very high-mass disks with $Q$-value less than
2 or around, the disk turbulence is important and the vertical
hydrostatic balance may not be reached.  

In order to see this self-gravity effects in the actual observations, 
it is necessary to have a very good sensitivity in the NIR direct
imaging observations.   
As we have seen in Figure \ref{fig:NIRpeak_Q}, it is necessary to detect
the peak with the strength of 5 to 10 percent when the disk $Q$-value is
about 3-4.  
The typical error of the NIR disk imaging observations is rather 
difficult to estimate, since it depends on how well the stellar light
can be suppressed.   
In the specific example of the observation of the disk around AB Aur by
AEOS telescope \citep{O08}, 
\citet{JCK10} estimate that the error is typically at the level of 
$\sim 20\%$ by
fitting the intrinsic azimuthal variation by a certain fitting function
and by calculating the rms deviation.  
It is above the level of the peak strength calculated in this paper.  
However, in future observations, better suppression of the speckle noise
may be made possible.  

For the particular model presented in this paper, the required
resolution of the telescope can be reached by the current instrument
like Subaru.  We have checked this by assuming the disk is at 140pc and
convolve the NIR profile with the Gaussian with FWHM $0.04^{''}$, 
which is the typical resolution of the Subaru telescope.  The resulting
image is not much different from the model image.  
$0.04^{''}$ at 140pc corresponds to 5.6$\mathrm{AU}$, but the model
image varies with the spatial scale of at least 10$\mathrm{AU}$.

We note that although we show that there may be a feature unique to
a self-gravitating disk in the NIR scattered light observations, it is
essential to compare this feature with the sub-mm observations.  
Since we can observe only the tenuous surface of the disk in the NIR
observations, small surface structure, such as a blob, can also cause
the bump-like structure which casts a shadow over the disk \citep{JCB07}.  
It is therefore necessary to have a sub-mm observation which
suggests that the disk is truly massive and there exists a gap in the
surface density.  
The correlation between NIR and sub-mm images must also be checked in
order to assure that the hydrostatic balance in the disk vertical
direction is reached.  

Sub-mm observations are also important in determining how the bump is
produced.  
Another mechanism that may cause the bump-like feature in the NIR
scattered light may be localized pressure bumps or a migrating
protoplanet.  A pressure bump may appear, for example, as a result of
baroclinic instability \citep[e.g., ][]{KB03}.  A protoplanet migrating 
inward in a disk may accumulate gas at the inner edge of a gap and
therefore, a bump-like feature may appear \citep[e.g., ][]{R02}.  
These processes both act even if the disk mass is low enough and the
effects of self-gravity is negligible.   
Since both processes involve the accumulation of the gaseous materials, 
such bump structures may also appear in the sub-mm dust continuum
emission.  It is therefore stressed again that the combination of the
sub-mm and the NIR observations is important.

The sub-mm imaging observations should also be used to make a disk model
that can be compared with NIR observations.  We note that the results
presented in Figure \ref{fig:NIRpeak_Q} are based on the specific model
used in this paper.  Although we expect that the general trend 
of the relation between the $Q$-value and the peak strength is not very
much different if we incorporate other forms of the gap profile or other
disk models, it is necessary to make a model that fits with the sub-mm
observation data of the surface density profile.   
It is essential to have at least the same level of spatial resolution
between the NIR and sub-mm telescopes.

In summary, if the disk mass is high and the $Q$-value is less than
$\sim 4$, the effects of self-gravity would come into play in
determining the disk vertical structure in hydrostatic equilibrium, 
and this effect would appear in future 
NIR direct imaging observations as a detailed structure such as a bump
and a gap, if we have sufficient sensitivity.  
The interpretation of the NIR
imaging data would therefore become complicated, 
especially in high-mass disk case.   
However, such detailed features may carry important information about
the physical states of the disk such as disk mass.
The combination between 
the sub-mm imaging observations and NIR imaging observations are also
important in the interpretation of the image.

\section{Summary and Caveats}
\label{sec:discuss}

In this paper, we have investigated the effects of self-gravity 
on the vertical structure of a protoplanetary disk with a gap, 
and discussed the possible signatures of a self-gravitating disk 
in direct imaging observations.
We have seen that the self-gravity of the disk causes a bump
feature to appear at the edge of the gap in the NIR imaging
observations, especially for the disks with $Q$-parameter 
with $\sim 3-4$.   
We have discussed that, 
although the interpretation of the NIR image would become complicated
due to the effects of self-gravity if the disk is very massive, 
detailed analyses of the morphology of the observed image would provide
us useful information about the physical state of the disk.  

The combination of the sub-mm and NIR images is important to assure
that the disk is truly massive and is in hydrostatic equilibrium 
in the vertical direction, 
and to check whether other possible causes of the bump may play a role.
Also, since the strength of the dust continuum emission gives a rough
estimate of the surface density, it is possible to check the consistency
of the model (i.e., $f_{\rm enhance}$ may be estimated both from the
peak strength and the strength of the dust continuum emission) if such
feature is detected.

Although we have considered a specific disk model 
with an axisymmetric gap, we expect that 
the bump structure associated with the gap may be a universal
feature in a self-gravitating disk that has a rapid variation in the
$Q$-value. 
This feature comes from the fact that the surface of the disk is at
the highest altitude in the vertical direction at a certain value of
$Q$-parameter.  The ring geometry is not essential in producing such a
bump associated with a gap. 

We note that we have assumed the vertical hydrostatic equilibrium, and
therefore, our model cannot be used for a turbulent disk.  The
turbulence induced by the self-gravity is expected to occur when the
disk $Q$-value becomes less than $\sim1.5-2$, and in this case, the
structure in the upper layer of the disk may be totally uncorrelated
with that at the midplane.  

One of the main caveats of the model presented in this paper is the
simplification of the temperature structure. 
Although the assumption of a vertically isothermal disk may not be 
seriously wrong for the outer disk, 
the irradiation from the central star can cause a super-heated layer at 
the surface of the disk \citep{CG97}, 
which can modify the observed signature in the NIR wavelength.  
We expect that the surface of the disk may be pumped up by the
irradiation from the central star, and the bump structure 
we have discussed in this paper is more enhanced.  
Therefore, we expect that the effects of 
irradiation may enhance the feature caused by the self-gravity.
However, this issue is necessary to be investigated separately.

\acknowledgments
The author of this paper thanks Munetake Momose, Akio K. Inoue, and
Misato Fukagawa for useful comments.  
He also thanks the anonymous referee for useful comments that improved
the paper.  
He is supported by Grants-in-Aid
for Japan Society for the Promotion of Science (JSPS) Fellows
(22$\cdot$2942). 

\appendix

\section{Grazing Angle}
\label{app:graze}

In this Appendix, we show the calculations of the grazing angle between
the disk surface and the incident ray in the case of
a non-self-gravitating disk.  
The objective of this section is to show explicitly that the
observed NIR flux should be well correlated 
with the surface density of the disk, 
which is proportional to the observed sub-mm flux, 
if the following two conditions are satisfied:
(1) the self-gravity of the disk is negligible 
and (2) the vertical hydrostatic balance is reached.  

If the hydrostatic equilibrium in the vertical direction is reached, the
density profile of the non-self-gravitating, vertically isothermal disk
is given by 
\begin{equation}
 \rho(r,z) = \rho_0(r) \exp
  \left[
   -\dfrac{z^2}{2H^2}
  \right],
\end{equation}
where $H$ is the pressure scale height and $\rho_0(r)$ is the density at
the midplane.  The midplane density is related to the surface density by
\begin{equation}
 \rho_0 (r) = \dfrac{\Sigma(r)}{\sqrt{2}H(r)}.
\end{equation}
We first calculate the optical depth, $\tau_{\rm rad}$, 
between the central star and a point $(r,z)$ in the disk.  Since
$\tau_{\rm rad}$ is calculated by
\begin{equation}
 \dfrac{d \tau_{\rm rad}}{dl} = \chi \rho,
\end{equation}
where $l$ is the path connecting between the point $(r,z)$ and the
central star, we have
\begin{equation}
 \tau_{\rm rad}(r,z) = \chi \dfrac{\sqrt{r^2+z^2}}{r}
  \int_{r_{\rm in}}^{r} dr^{\prime}
  \rho_0(r^{\prime}) 
  \exp \left[  
	-\dfrac{1}{2h(r^{\prime})^2} \dfrac{z^2}{r^2}
       \right],
  \label{taurad}
\end{equation}
where $h=H/r$ and we have assumed that the central star is a point
source situated at the origin.

We first look at the approximate position of the disk surface.  We
define the disk surface as a place where $\tau_{\rm rad}=\tau_s=1$.  
We assume that $z \ll r$, which is justified later.   
The estimate of the order of magnitude of the integration in equation
\eqref{taurad} yields
\begin{equation}
  \int dr^{\prime}
  \rho_0(r^{\prime}) 
  \exp \left[  
	-\dfrac{1}{2h(r^{\prime})^2} \dfrac{z^2}{r^2}
       \right],
  \sim \rho_0 r e^{-z^2/2H^2}
\end{equation}
and therefore, $\tau_{\rm rad}$ is
\begin{equation}
 \tau_{\rm rad} \sim \chi \rho_0 r e^{-z^2/2H^2}.
\end{equation}
Therefore, the surface $z_s$ of the disk may be calculated by
\begin{equation}
 z_s \sim H \sqrt{ 2 \log \left( \dfrac{\chi \rho_0 r}{\tau_s} \right) }
  \sim H 
  \sqrt{ 2 \log \left( \dfrac{\chi}{\tau_s} 
		 \dfrac{\Sigma_0}{\sqrt{2}h}  \right)}.
  \label{surf_approx}
\end{equation}
If we use 
$\Sigma_0 = 10 \mathrm{g/cm}^2$, 
$\chi=100\mathrm{cm}^2\mathrm{/g}$, 
$\tau_s=1$ and $h=0.1$, we have $z_s\sim 4H$.  
Therefore, the disk surface resides at several scale heights above
the disk midplane.  Since $H \ll r$ in a protoplanetary disk, this
confirms the initial assumption of $z \ll r$. 

We now look at the grazing angle $\beta$ between the disk surface and
the incident ray.  Since $z_s \ll r$, $\beta$ can be calculated, to a
good approximation, by
\begin{equation}
 \beta \sim \dfrac{dz}{dr}\Bigg{|}_{\mathrm{surf}} 
  - \dfrac{z_s}{r}, 
\end{equation}
where $dz/dr$ should be evaluated along the disk surface,
\begin{equation}
 \dfrac{dz}{dr}\Bigg{|}_{\mathrm{surf}}
  = -\dfrac{\partial \tau_{\rm rad}/\partial r}
  {\partial \tau_{\rm rad}/\partial z}
\end{equation}  
It is to be noted that from equation \eqref{taurad}, 
$\tau_{\rm rad}(r,z)$ is a function of $r$ and $z/r$ only.  The
grazing angle $\beta$ is calculated as
\begin{equation}
 \beta = - \dfrac{\sqrt{1+(z/r)^2} \rho_0(r) \exp[(1/2h^2)(z/r)^2]}
  {\partial_u \left[ \sqrt{1+u^2}\int^r dr^{\prime} \rho_0(r^{\prime}) 
	      \exp[-u^2/2h^2]\right]\Big{|}_{u=z_s/r} }.
\end{equation}
Since $z_s/r \sim h \ll 1$, the denominator can be approximated as
\begin{equation}
 -\dfrac{1+(z_s/r)^2}{z_s/r} \int^r dr^{\prime}
  \rho_0(r^{\prime}) \dfrac{(z_s/r)^2}{h(r^{\prime})^2}
  e^{-u^2/2h(r^{\prime})^2}.
\end{equation}
Therefore, the grazing angle is given by
\begin{equation}
 \beta \sim \dfrac{z_s}{r}
  \dfrac{ \rho_0(r) \exp[-(1/2h^2)(z_s/r)^2]}
  {r^{-1}\int^r dr^{\prime} \rho_0(r^{\prime}) (z_s/r h(r^\prime))^2 
  \exp[-1/2h(r^{\prime})^2 (z_s/r)^2]}.
  \label{beta_approx}
\end{equation}
Since $z_s/r \sim h$, the order of the magnitude of $\beta$ may be
\begin{equation}
 \beta \sim h \dfrac{\rho_0}{ (1/r) \int dr^{\prime} \rho_0(r^{\prime})}.
\end{equation}
Therefore, we conclude that 
if there exists a gap in the disk and the gap width is narrow compared
to the disk scale, the value of $\beta$ is almost proportional to the
midplane density, which is proportional to the surface density since we
assume the hydrostatic equilibrium in a non-self-gravitating disk.
Therefore, in this case, the NIR scattered light, whose intensity is
proportional to the grazing angle, should be well-correlated with the
sub-mm thermal emission, which reflects the profile of surface density.

\clearpage

\end{document}